# Terms in journal articles associating with high quality: Can qualitative research be world-leading?

Mike Thelwall
Statistical Cybermetrics and Research Evaluation Group, University of Wolverhampton, UK.
https://orcid.org/0000-0001-6065-205X m.thelwall@wlv.ac.uk

Kayvan Kousha
Statistical Cybermetrics and Research Evaluation Group, University of Wolverhampton, UK.
https://orcid.org/0000-0003-4827-971X k.kousha@wlv.ac.uk

Mahshid Abdoli
Statistical Cybermetrics and Research Evaluation Group, University of Wolverhampton, UK.
https://orcid.org/0000-0001-9251-5391 m.abdoli@wlv.ac.uk

Emma Stuart
Statistical Cybermetrics and Research Evaluation Group, University of Wolverhampton, UK.
https://orcid.org/0000-0003-4807-7659 emma.stuart@wlv.ac.uk

Meiko Makita
Statistical Cybermetrics and Research Evaluation Group, University of Wolverhampton, UK.
https://orcid.org/0000-0002-2284-0161 meikomakita@wlv.ac.uk

Paul Wilson
Statistical Cybermetrics and Research Evaluation Group, University of Wolverhampton, UK.
https://orcid.org/0000-0002-1265-543X pauljwilson@wlv.ac.uk

Jonathan Levitt
Statistical Cybermetrics and Research Evaluation Group, University of Wolverhampton, UK.
https://orcid.org/0000-0002-4386-3813 j.m.levitt@wlv.ac.uk

**Purpose**: Scholars often aim to conduct high quality research and their success is judged primarily by peer reviewers. Research quality is difficult for either group to identify, however, and misunderstandings can reduce the efficiency of the scientific enterprise. In response, we use a novel term association strategy to seek quantitative evidence of aspects of research that associate with high or low quality.
**Design/methodology/approach**: We extracted the words and 2–5-word phrases most strongly associating with different quality scores in each of 34 Units of Assessment (UoAs) in the Research Excellence Framework (REF) 2021. We extracted the terms from 122,331 journal articles 2014-2020 with individual REF2021 quality scores.
**Findings**: The terms associating with high- or low-quality scores vary between fields but relate to writing styles, methods, and topics. We show that the first-person writing style strongly associates with higher quality research in many areas because it is the norm for a set of large prestigious journals. We found methods and topics that associate with both high- and low-quality scores. Worryingly, terms associating with educational and qualitative research attract lower quality scores in multiple areas. REF experts may rarely give high scores to qualitative or educational research because the authors tend to be less competent, because it is harder to make world leading research with these themes, or because they do not value them.
**Originality**: This is the first investigation of journal article terms associating with research quality.

# Introduction

Academic research in increasingly many countries is evaluated by post-publication peer review for quality assurance, formative feedback or to direct research funding (Engels & Guns, 2018; Woelert & McKenzie, 2018; Sivertsen, 2018; Jeon & Kim, 2018; Nielsen, 2017). The results can influence the reputations, funding, actions, and careers of the researchers involved. It is therefore important to investigate any systematic causes of high- or low-quality scores to find areas of good or bad practice and to check for bias. This is inherently difficult for academic peer review because even experts can strongly disagree on what is good research. Thus, whilst quality differences or biases can be tested for in researcher characteristics (e.g., gender, career status) or institutional status (e.g., prestige, size, geographic location), it is difficult to identify content-related patterns, such as research topics or methods that tend to attract high or low scores. The primary difficulty is that each research article is unique and not flagged with its core characteristics. Perhaps the most relevant general data for an article is its set of keywords, but not all articles have these, authors use them differently, and controlled vocabularies are not universal. In response, we import and adapt a social science word frequency analysis method, word association contextualisation (Thelwall, 2021), and apply it to the titles, abstracts and keywords of the articles evaluated to explore for article content characteristics that associate with high or low research quality.

Research quality is usually characterised as combining rigour, originality, and societal/academic significance (Langfeldt et al., 2020; REF2021, 2020). Whilst rigour is relatively objective, originality is subjective (Sánchez et al., 2019) and all three components depend on the expertise of the evaluator. For example, a statistical expert might be more critical of the quantitative component of the methods but might not notice the unusual degree of care that a qualitative researcher has taken to safeguard participant safety. More generally, expert reviewers can be the most critical (Gallo et al., 2016). Reviewers might also be biased by gender (Ceci & Williams, 2011; Fox & Paine, 2019), nationality (Harris et al., 2019; Primack et al., 2009; Thelwall et al., 2021), ethnicity (Woolston, 2021), and prestige (Tomkins et al., 2017). Cognitive cronyism, in the sense of judging results from known specialism better, is widely suspected but with little evidence (Lee et al., 2013; Wang & Sandström, 2015), and it is possible that cognitive cronies are more critical because they are more expert (e.g., Gallo et al., 2016). Indirect support for the hypothesis can be found from evidence that academics are more impressed by journals from their own specialties than others (Serenko & Bontis, 2018). Academics tend to give lower quality ratings to articles with conclusions that conflict with their beliefs, at least in psychology (Hergovich et al., 2010), perhaps because they are more suspicious of them. All these factors might explain the low degree of agreement between peer reviewers in many contexts (Fogelholm et al., 2012; Jackson et al., 2011; Kravitz et al., 2010; Pier et al., 2018; Rothwell et al., 2000; c.f., Pina et al., 2015), but do not suggest any content-related factors that might partly determine the quality of published research.

Some content-based factors are also known to relate to the citation impact or quality of an article. Findings might be judged to be more important if they are positive or statistically significant (Easterbrook, et al., 1991; van Lent et al., 2014). Individual research methods associate with differing levels of average citation impact, which may relate to their quality. In particular, articles reporting interviews, case studies, focus groups and ethnographies tend to be less cited in most fields (Thelwall & Nevill, 2021). Conversely, research mentioning questionnaires (Fairclough & Thelwall, 2022) or structural equation modelling (Thelwall & Wilson, 2016) tends to be more cited. There are field-based exceptions to these trends, however (no citation difference in library and information studies: Jamali, 2018; a qualitative advantage in international business research, although it classed survey articles as qualitative: López-Morales et al., 2022). Within most fields there are probably highly cited topics (e.g., Kim et al., 2011; Sanchez, 2020; Savin & van den Bergh, 2021), although it is not clear whether such topics would be generally agreed to include above average quality research. Interdisciplinary research may receive lower scores if the evaluators expect it to meet all the quality criteria of its constituent fields, so discussion between reviewers is helpful to understand the work from a holistic perspective (Huutoniemi, 2012; Oviedo-García, 2016). There are also legitimate types of methods bias in quality assessments based on hierarchies of evidence in some health-related fields (Katz et al., 2019; Murad, et al., 2016). For example, other factors being equal, a double-blind placebo-controlled randomised control trial is methodologically far more rigorous than a professional opinion (Vere & Gibson, 2021), although it could be considered less original.

No prior study has explored which contents of articles associate with research quality rather than citation impact. Terms in article titles, abstracts and/or keywords are often used to map research topics (e.g., Ravikumar et al., 2015). Moreover, keywords (Kim et al., 2011) and manually identified themes (Sanchez, 2020) have been checked for associations with citation impact. The frequency of words or two- or three-word phrases in titles, abstracts, and keywords of articles within Scopus narrow categories has also been used to find terms occurring associating with hot topics showing them to be more cited in most fields (Thelwall & Sud, 2021). Repetition of keywords in abstracts also associates with citation counts for education journals (Sohrabi & Iraj, 2017) and the presence popular management information system keywords can more effectively predict highly cited papers (n=746) than journal (e.g., Journal Impact Factor and SCImago Journal Rank) or author (author's h-index, publications, or citations) features (Hu et al., 2020).

In response to the scarcity of general evidence of the relationship between research quality and article contents, we applied a modified version of word association contextualisation to detect words and themes associated with high- or low-quality research, as evaluated in REF2021. This is an exploratory method in the sense that we tested no hypotheses. Instead, the method itself generates the words and themes that are its output. The following general research questions drive our analysis.

- RQ1: Which types of words or phrases in articles and titles associate with higher quality research, if any?
- RQ2: Does the answer to RQ1 vary between fields?
- RQ3: Do the answers to RQ1 and RQ2 have wider implications for research evaluation?

## Methods

We used 148,977 journal articles submitted to REF2021 by publicly funded higher education institutions, except the University of Wolverhampton, which were redacted. Each article had identifying information, such as the journal, title and DOI, as well as its provisional score, as of March 2022. The results were published in May 2022 and the provisional results are very close to the final values, according to the REF team that supplied them. For confidentiality reasons, we deleted the data on May 8, 2022.

We removed 318 unclassified articles before the analysis. The REF records did not contain article titles abstracts and keywords, so we matched the articles to Scopus 2014-2020 for these. We searched REF outputs by DOI in a local copy of Scopus, generating 133,218 matches. We automatically searched the remaining articles by title and journal name (after converting to lower case and removing spaces), and manually checked the results to filter out false matches (typically articles with sort generic titles, such as, "Comment") to give an additional 997 results. We removed additional copies of articles that had been submitted by multiple institutions to the same UoA (or Main Panel for the panel-based analysis). For duplicate articles in the same UoA or Main Panel, we used the median score or one of the two medians at random when there was a tie. We analysed the articles extracted primarily by UoA to give the finest grained results, with UoAs grouped into four Main Panels and one complete set to identify more general trends. We removed articles scoring 0 since these or their authors may have been out of scope. We also removed articles with abstracts shorter than 500 characters, after cleaning, because these seemed to be a different type of output, such as a letter or comment, and therefore not comparable (Table 1).

Table 1. The number of journal articles submitted to the REF and matching Scopus 2014-2020, after removing duplicates and removing articles with cleaned abstracts shorter than 500 characters.

| UoA or Panel | Articles |
|---|---|
| 1 | 9905 |
| 2 | 3889 |
| 3 | 9675 |
| 4 | 8172 |
| 5 | 6376 |
| 6 | 3147 |
| 7 | 3724 |
| 8 | 3274 |
| 9 | 4499 |
| 10 | 5111 |
| 11 | 4645 |
| 12 | 16333 |
| 13 | 2582 |
| 14 | 3439 |
| 15 | 545 |
| 16 | 1762 |
| 17 | 11851 |
| 18 | 1864 |
| 19 | 2502 |
| 20 | 3294 |
| 21 | 1498 |
| 22 | 977 |
| 23 | 3336 |
| 24 | 2812 |
| 25 | 524 |
| 26 | 962 |
| 27 | 768 |
| 28 | 1082 |
| 29 | 111 |
| 30 | 806 |
| 31 | 185 |
| 32 | 1117 |
| 33 | 544 |
| 34 | 1020 |
| Main Panel A | 37282 |
| Main Panel B | 36584 |
| Main Panel C | 35631 |
| Main Panel D | 7071 |
| **UoA total** | **122331** |
| **Panel total** | **116568** |

We used chi square tests to identify words that occurred disproportionately often in articles with different quality levels. First, we cleaned article abstracts for journal standard texts, such as copyright statements, structured headings, and open access statements. This cleaning was automatic using a large set of heuristics initially created for a previous study (Thelwall & Sud, 2021) and updated for our programme of work on the REF outputs. After the data cleaning, we extracted words and phrases of up to three words not spanning sentence boundaries from each article. A limit of five words in a phrase was set to capture relevant short phrases without overwhelming the results with longer, over-specific terms. We merged the lowest two scores (1* and 2*) into a single group to increase statistical power since the 1* group was very small.

In each UoA and Main Panel, we calculated a chi square value for every word and phrase extracted to assess whether it occurred disproportionately often in one of the three quality groups (1* or 2*, 3*, 4*). Value reflects whether the three quality classes have different proportions of articles containing each term. For example, if 1% of 1* or 2* articles contained "funded by", 2% of 3* articles contained "funded by" and 5% of 4* articles contained "funded by" then these differences would translate into a large chi square value, and if the percentages were identical (e.g., all 2%) then the chi square value would be 0.

For each UoA and main panel, we examined the fifty terms with the highest chi squared values and grouped them into three themes according to their main apparent purpose. The three themes we found were style, methods, and topic. The first author performed the classifications. Since spurious statistical positives can occur with multiple significance tests, a Bonferroni correction (Ranstam, 2016) was applied to identify a minimum chi square value (i.e., a corrected α=0.05 significance level) in each UoA or Main Panel for a term to be statistically significant. We ignored terms with a chi square below this value, even if they were in the top 50 for the UoA or Main Panel. In cases where this resulted in no statistically significant terms for a UoA, we added terms with the highest chi square values for illustrative purposes and flagged them as such.

# Results

## *Main Panel A: Words and phrases*

For Main Panel A, which mostly focuses on life sciences, health and medicine, there are many stylistic terms that associate with prestigious journals (Table 2). In particular, "here we show that" is a common phrase in enough prestigious journals in this area to statistically associate with high quality research. This phrase occurs in abstracts, usually in the middle after introducing the context of the study, but sometimes at the start. More generally, first person plural singular (we, our) in the present tense is a common style in several prestigious journals in this area (e.g., *Blood, Cell*), whereas the third person and past tense more associate with other journals (e.g., "this study was"). For example, whilst 60.8% of REF journal article extracts contained "we", it was in 99% of abstracts in the journal *Nature,* 96% of *Science,* 94% of *The Lancet,* and 91% of *Cell*. This suggests that it is an actual or de facto style requirement for these journals.

Funding associated with higher quality in most UoAs in this Main Panel, typically through a declaration at the end of an abstract, such as, "Funding: This project was funded by the NIHR." This seems to be a journal style issue because funding was mainly mentioned *The Lancet* and its family of journals (e.g., *The Lancet Public Health*). It is unlikely to be a funding issue since most studies in this area were presumably externally funded.

Methods terms partly reflect standard hierarchies of evidence or at least indicators of higher quality studies (e.g., double-blind, masked, "randomly-assigned patients"). In some UoAs, qualitative methods associate with lower quality (e.g., interviews, themes, thematic, qualitative) although the term "measured" also associated with lower quality in one. The suggestion that qualitative research tended to score lower could be due to bias in favour of quantitative research or adherence to evidence hierarchies that do not include qualitative studies. Based on reading abstracts in UoA 3, containing "themes" or "thematic", since qualitative studies are based on limited samples, they provide insights and may trigger suggested actions (e.g., "Resources are needed that are tailored to men, framed around fatherhood") but it might be more difficult to argue that the findings are world leading because the conclusions seem unlikely to have direct societal impact or to be definitive. This is possibly a generic issue with evaluating qualitative research.

In some UoAs there were terms indicating topics that tended to be higher or lower quality, although there is also an overlap with methods. For example, mice and mouse models are used in many research methods. The topics may associate with the main areas of strong or weak research groups submitted to each UoA rather than being intrinsically more important.

Table 2. Examples of words and phrases with the strongest associations (chi-square test) with REF scores by UoA for Main Panel A. Bold terms associate with lower REF scores; other terms associate with higher REF scores. Words and shorter phrases within longer phrases are omitted in favour of the longest relevant phrase.

| UoA | Style | Methods | Topic |
|---|---|---|---|
| 1: Clinical Medicine | We, here, **were** | Funding, "randomly assigned patients", "the primary outcome", double-blind, interpretation, masked, "trial is registered with", "in the placebo group", "group and", "to receive", "primary outcome", intention-to-treat, "adverse events" | |
| 2: Public Health, Health Services and Primary Care | "This is an" | Funding, "randomly assigned", "the primary outcome", randomisation, interpretation, trial, "trial is registered", "adverse events", "group and", **interviews, participation** | |
| 3: Allied Health Professions, Dentistry, Nursing and Pharmacy | "Here we", "we show that", **was, were** | Funding, CI, "trial is registered", "adverse events", randomised, randomisation, "randomly assigned", "the primary outcome was", intention-to-treat, stratified, "adverse events", **themes, thematic** | |
| **4: Psychology, Psychiatry and Neuroscience** | "Here we show that", **were, was, "the aim", "the current study", "the aim"** | Funding, "randomly assigned", "trial is registered", "the primary outcome was", vivo, **online, web, "measures of", completed, discussed, research, qualitative, interviews** | Neurons, neuronal, gene, human, mouse, cell, protein, brain, synaptic, disease, **participants** |
| 5: Biological Sciences | "Here we show that", **be, were, "this study was", "did not", no, investigated, conducted, some, "of this study", "used to", may, had, been, or** | Web, "the effects of", "a significant", compared, assessed, "to assess", mean, measured | |
| 6: Agriculture, Food and Veterinary Sciences | "Here we show that", "we report", our, **were, "there was", significantly, "used to", on, no, had, "this study", "this paper", be,** | "We identify", replication, **collected, "evaluation of", "the effect of"** | "Evolution of", signaling, evolutionary, genes, cells, "Amino acid", mice, genome, genomic, "in arabidopsis", mechanism, **horses, "in dogs", "dogs with", farm Cells** |
| **Panel A** | "Here we show that", **were, was, "this study"** | Funding, "we randomly assigned", "is registered with", "the primary outcome was", interpretation, "adverse events", "to receive" | |

## Main Panel B: words and phrases

For Main Panel B (Table 3), there were similar journal style terms to Main Panel A. For methods, the results suggest that experimental work and proof tended to be rated higher quality and that qualitative research (again) tended to attract lower scores. Case studies are also mentioned for Main Panel B overall. This term could refer to the case study method or an investigation of a single example of something using other methods. The topics for Main Panel B UoAs could reflect both research group specialisms and important societal topics (e.g., warming, climate, ocean). Research mentioning students or higher education tended to be lower quality.

Table 3. Examples of words and phrases with the highest chi-square values by UoA for Main Panel B. Bold terms associate with lower scores; other terms associate with higher scores.

| UoA | Style | Methods | Topic |
|---|---|---|---|
| 7: Earth Systems and Environmental Sciences | "Here we show that", "we find that", "here we present", **"was also", were, "in this study", showed** | Analysis, significant, investigated, method, behaviour, "compared to" | "The global", warming, earth, climate, ocean, "million years", "years ago", atmospheric, ice, circulation, forcing, **UK Raw, wort** |
| 8: Chemistry | "Here we", "we show that", **"were performed", was, showed, "were found to", "an investigation into"** | Reduce, tests, assessment, evaluated, "the formulations" "and in vitro" | |
| 9: Physics | "Here we report", "so far", | "Has been developed" | |
| 10: Mathematical Sciences | We | "We prove", **visualisations, recruited, "is in the use of", "trial registration", "and simplify the"** | **"Only a subset", "codes over rings", epidemiology, "the epidemic", aged, "from group rings"** |
| 11: Computer Science and Informatics | "We show that", "we demonstrate that", "we introduce", our, **"this study", "the results", "this research", "this paper", project, "the results", presented** | Experiments, approximate, complexity, "we prove that", bounds, probabilistic, **review, interviews, development** | Problem, imaging, general, graph, first, polynomial, "class of", **technology, future** |
| 12: Engineering | "Here we report/present/demonstrate", "here we show that", **were, was, had, study, "results showed", "this paper", "the results", investigated, "carried out"** | "In vivo", **qualitative, interviews, analysis** | Imaging, photonic, quantum, optical, spatial |
| Panel B | "Here we show that", "here we demonstrate", "here we present", "we report", our, **was, were, "the purpose of this paper", "this study", "this research", used, "the results", showed, investigated, "the proposed", different** | "We prove", **analysis, "case study", compared, performance** | Warming, global, "years ago", "million years", quantum, **management, students, "higher education"**. |

## Main Panel C: words and phrases

For Main Panel C, there were again style terms with a tendency for the first-person present tense to be higher quality and third person past tense to be lower quality (Table 4). This was again primarily journal-based. For example, in UoA 13 only 9 out of 100 *Energy and Buildings* article abstracts contained "we" but it was in 39 out of 40 *Nature* family journal article abstracts. Qualitative methods were again given lower quality scores overall and for one UoA. Important global issues again scored well and, more clearly than before, education-related articles tended to attract lower scores (although not in UoA 23).

Table 4. Examples of words and phrases with the highest chi-square values by UoA for Main Panel C. Bold terms associate with lower scores; other terms associate with higher scores.

| UoA | Style | Methods | Topic |
|---|---|---|---|
| 13: Architecture, Built Environment and Planning | **"The purpose of this paper"** | | |
| 14: Geography and Environmental Studies | "Here we show", "we show that", "our results", "we find that", **"of this study"**, **"the study"**, **"the results"**, were, "there was a", used, showed | "Per cent" | "In global", "a global", climate, oceanic, tropical, "earth system", **UK** |
| 15: Archaeology | "Here we", **"this article"**, | | Asia |
| 16: Economics and Econometrics | We, "we develop", indicate, **"purpose of this paper is"**, **"the authors"**, **"based on an"** "this study", "the findings", "willing to" | | **Annual**, **"for future"**, region |
| 17: Business and Management Studies | "We show", "we develop", when, "we find that", **"purpose of this paper is"**, **"of this paper is to"**, **"have been"**, **was"**, **"the findings"**, **there** | "consistent with", **review**, **analysis** | Behaviour, **"the UK"**, **"the period"**, policy, sector, crisis |
| 18: Law* | "We argue that" | | **Students** |
| 19: Politics and International Studies | "We find that", our, "we show that", **also**, **"this article"** | Data, effects, results, experiment | Electoral |
| 20: Social Work and Social Policy | "We find that", "we use", our, show, **"the purpose of"**, **researcher**, **research** | "Longitudinal study", long-term, panel, CI, "data for", cohort, effects, evidence, results, estimate, per, baseline, rates, modelling, **themes** | Household, birth, family, incentives, "the English", at age, **students**, **teaching**, **learning** |
| 21: Sociology* | We | Models, "longitudinal study" | |
| 22: Anthropology and Development Studies* | "We present" | Consistent | Modalities |
| 23: Education | "We find" | Longitudinal, multilevel, measures, **"perceptions of"**, **experiences** | Attainment, **staff**, **online** |
| 24: Sport and Exercise Sciences, Leisure and Tourism | "Here we", "we show", **"this study"** | "Muscle biopsies were", "muscle protein synthesis", **"in vivo"**, **"total distance"**, **completed** | "Human skeletal muscle", "of muscle", "muscle mass", humans, expression, motor, atrophy, **"countermovement jump"**, **"soccer players"**, **sport** |
| Panel C | "we show that", "we find that", "here we", our, **"the purpose of this paper"**, **"aims to"**, **"the research"**, **"this study"**, **challenges** | "Consistent with", effects, **interviews**, **themes** | "Skeletal muscle", **"higher education"**, **university**, **students**, **teaching**, **teachers**, **staff**, **experiences**, **issues**, **development** |

* Listed terms are not statistically significant after a Bonferroni correction.

## *Main Panel D: words and phrases*

No Main Panel D UoA had any statistically significant terms, but students are gain mentioned overall as a lower quality topic (Table 5).

Table 5. Examples of words and phrases with the highest chi-square values by UoA for Main Panel D. Bold terms associate with lower scores; other terms associate with higher scores.

| UoA | Style | Methods | Topic |
|---|---|---|---|
| 25: Area Studies* | | | "The global", governments |
| 26: Modern Languages and Linguistics* | | Results | Semantics, **narrative** |
| 27: English Language and Literature* | Whose | | Phonetic, **"of lexical"** |
| 28: History* | **Will** | | "The old" |
| 29: Classics* | | Concept, "reference to" | Wider |
| 30: Philosophy* | "It is" | Statistical, **"the historical"** | |
| 31: Theology and Religious Studies* | "In some" | Associated | "The phenomenon" |
| 32: Art and Design: History, Practice and Theory* | "This essay" | Examination | "And artistic" |
| 33: Music, Drama, Dance, Performing Arts, Film and Screen Studies* | **My** | Rated | Musical, music, **performance** |
| 34: Communication, Cultural and Media Studies, Library and Information Management* | Most | | Senior, move, **"the local"** |
| Main Panel D | | | Syntactic, variation, **students, narrative** |

* Listed terms are not statistically significant after a Bonferroni correction.

## Discussion

The limitations of the results include that the journal articles all have at least one UK author, and are self-selected, with a cap of 5 per researcher. This may help researchers in quantitative subjects that produce more work and can cherry pick their best outputs. Conversely, it may also help researchers that produce less work because each submitted output represents a larger share of their efforts. The REF system may also have pushed people that are primarily educators into conducting education-related research to participate, where they would compete with people that devote more time to research. If true, then the education research may have more value per quality point since it would have had less input. Moreover, the REF2021 rules may have affected the results as may any discussions within UoAs or main panels about the relative merits of different types of research. The results also rely on what is written in titles abstracts and keywords, which may not translate directly to the topics of articles. For example, perhaps researchers that are more expert with qualitative research use more specific terms than "theme" or "qualitative", giving a second order quality effect for the remaining articles using these terms.

The stylistic results are relatively uninteresting in the sense that they point to journal-based norms and differences in the average quality of journals are well known. It is therefore unsurprising that journal style norms translate into quality-associated stylistic terms. It is also possible that higher quality articles in other journals have similar stylistic features, either because the authors are experienced in submitting to prestigious journals, associate the style with high quality research, or publish their best articles rejected from a prestigious journal to another type.

Some of the methods results for Main Panel A align with known hierarchies of evidence and good practice in medical fields by mentioning placebos, randomisation, double-blind, and trial registration (Vere & Gibson, 2021). Many of the other methods are quite specific and may relate to additional care taken with experiments or journal style guidelines about what to mention. The two main general results – the lower scores given to education-related research and qualitative research are different however, and do not seem to have been previously noted in studies of peer review bias. Nevertheless, there have been previous claims of general bias against qualitative research (Bansal & Corley, 2011) and those relying on hierarchies of evidence might regard it as being a low-level type (Vere & Gibson, 2021). Moreover, quantitative researchers used to larger sample sizes are known to sometimes devalue qualitative research for having few participants (Baillie & Douglas, 2014) or for lacking generalisability (Smith, 2018). Moreover, interviews and case studies tend to be less cited (Thelwall & Nevill, 2021). No previous study seems to have remarked that educational research tends to get lower quality scores in research evaluation contexts, however, although concerns have been raised that the distinction between educational research and teaching and learning scholarship is blurred (Cotton et al., 2018), which may have resulted in some sub-standard REF submissions, or suspicion on the part of assessors. In addition, the capability of REF assessors for educational research has been questioned (Cotton et al., 2018).

## Conclusion

The results show that there are stylistic methods and topic associations with different research quality scores for journal articles in most UoAs, especially those with large numbers of articles. Despite the focus on quality, no term directly mentioned an aspect of quality, such as through a claim to be novel, rigorous or impactful. The results suggests that there are common methodological associations with high scores, presumably because there are recognised hierarchies of method rigour. Since the style findings are journal-related and the individual topics could be due to individual research groups, the methods differences are the clearest general finding. Thus, a take-away message for researchers is to ensure that the most rigorous method is selected for each study.

The most worrying findings are the lower scores given in some UoAs to educational and qualitative research. As argued above, the former may be a systemic effect of the evaluation system; the latter is more concerning, given the need for methods pluralism in a healthy research system. This is particularly important for the REF where, at the time of writing, only 4* research was fully funded, with 3* receiving 25% funding and the remainder nothing. At the moment, interdisciplinary research is given special consideration within the REF rules. Special consideration may also be needed for qualitative research to ensure that it is not undervalued in academia.

## Acknowledgement

This study was funded by Research England, Scottish Funding Council, Higher Education Funding Council for Wales, and Department for the Economy, Northern Ireland as part of the Future Research Assessment Programme (https://www.jisc.ac.uk/future-research-assessment-programme). The content is solely the responsibility of the authors and does not necessarily represent the official views of the funders.